\documentclass[zpreprint,zbstpl,british]{zeus_Kpaper}
\usepackage{zeusdet_units}
\chardef\usc=95
\chardef\til=126
\catcode`\@=11 
\DeclareRobustCommand\xdotspace{\futurelet\@let@token\@xdotspace}
\def\@xdotspace{%
  \ifx\@let@token.\else
  \ifx\@let@token\bgroup.\else
  \ifx\@let@token\egroup.\else
  \ifx\@let@token\/.\else
  \ifx\@let@token\ .\else
  \ifx\@let@token~.\else
  \ifx\@let@token!.\else
  \ifx\@let@token,.\else
  \ifx\@let@token:.\else
  \ifx\@let@token;.\else
  \ifx\@let@token?.\else
  \ifx\@let@token/.\else
  \ifx\@let@token'.\else
  \ifx\@let@token).\else
  \ifx\@let@token-.\else
  \ifx\@let@token\@xobeysp.\else
  \ifx\@let@token\space.\else
  \ifx\@let@token\@sptoken.\else
   .\space
   \fi\fi\fi\fi\fi\fi\fi\fi\fi\fi\fi\fi\fi\fi\fi\fi\fi\fi}
\catcode`\@=12 
\newcommand{\CL}[1]{$#1\%$~C.L\xdotspace}
\newcommand{\stru}[2]{%
   \relax\ifmmode\hbox{\vrule height#1 depth#2 width0pt}%
   \else\vrule height#1 depth#2 width0pt\fi}

\newcommand{\Ronum}[1]{\uppercase\expandafter{\romannumeral#1}}
\newcommand{\ronum}[1]{\expandafter{\romannumeral#1}}
\DeclareRobustCommand{\LaTeXZ}{%
  \LaTeX\kern-.05em4\kern-.1em
  {\raisebox{-0.2ex}{$\scriptstyle\text{ZEUS}$}}\xspace}

\newcommand{\fig}[1]{Fig.~\ref{fig-#1}}

\newcommand{\slashfrac}[2]{%
  \raisebox{0.5ex}{\ensuremath #1}\kern-0.12em/\kern-0.08em
  \raisebox{-.8ex}{\ensuremath #2}}

\newcommand{\sqr}[3]{%
    {\vcenter{\hrule height.#3ex\hbox{\vrule width.#2ex height#1ex
     \kern#1ex\vrule width.#3ex}\hrule height.#2ex}}}

\catcode`\@=11 
\newcommand{\parenbar}{\mathpalette\p@renb@r}
\def\p@renb@r#1#2{\vbox{%
  \ifx#1\scriptscriptstyle \dimen@.7em\dimen@ii.2em\else
  \ifx#1\scriptstyle \dimen@.8em\dimen@ii.25em\else
  \dimen@1em\dimen@ii.4em\fi\fi \offinterlineskip
  \ialign{\hfill##\hfill\cr
    \vbox{\hrule width\dimen@ii}\cr
    \noalign{\vskip-.3ex}%
    \hbox to\dimen@{$\mathchar300\hfil\mathchar301$}\cr
    \noalign{\vskip-.3ex}%
    $#1#2$\cr}}}
\catcode`\@=12 

\ifzeusunit
  
\else
  
\fi

\newcommand{\IP}{{\rm I$\kern-0.01667em$P}\xspace}

\mathchardef\qsm=63
\mathchardef\pls=43
\mathchardef\mns=512
\mathchardef\plm=518
\mathchardef\eql=61
\mathchardef\smallleft=300
\mathchardef\smallright=301
\mathchardef\les=316
\mathchardef\gre=318
\mathchardef\leq=532
\mathchardef\grq=533

\catcode`\@=11 
\newcounter{pict@width}
\newcounter{pict@height}
\newlength{\pict@scale}
\newcommand{\psfigadd}[4]{%
\setcounter{pict@width}{1*\ratio{#2+\pict@scale/2}{\pict@scale}}
\setcounter{pict@height}{1*\ratio{#3+\pict@scale/2}{\pict@scale}}
\setlength{\unitlength}{\pict@scale}
\hbox to #2{\hspace{-\fill}\begin{picture}(\thepict@width,\thepict@height)
\put(0,0){\psfig{figure=#1,width=#2,height=#3,clip=}}
\SetScale{0.283466457}
\SetWidth{1.763889}
{#4}
\end{picture}}
}
\newcounter{pict@widthfst}
\newcounter{pict@widthscd}
\newcounter{pict@widthtot}
\newcommand{\psfigaddtwo}[7]{%
\setcounter{pict@widthfst}{1*\ratio{#2+\pict@scale/2}{\pict@scale}}
\setcounter{pict@widthscd}{1*\ratio{#2+#4+\pict@scale/2}{\pict@scale}}
\setcounter{pict@widthtot}{1*\ratio{#2+#4+#6+\pict@scale/2}{\pict@scale}}
\setcounter{pict@height}{1*\ratio{#3+\pict@scale/2}{\pict@scale}}
\setlength{\unitlength}{\pict@scale}
\hbox{\hspace{-\fill}\begin{picture}(\thepict@widthtot,\thepict@height)
\put(0,0){\psfig{figure=#1,width=#2,height=#3,clip=}}
\put(\thepict@widthscd,0){\psfig{figure=#5,width=#6,height=#3,clip=}}
\SetScale{0.283466457}
\SetWidth{1.763889}
{#7}
\end{picture}}
}
\newcommand{\psfigror}[4]{%
\setcounter{pict@width}{1*\ratio{#2+\pict@scale/2}{\pict@scale}}
\setcounter{pict@height}{1*\ratio{#3+\pict@scale/2}{\pict@scale}}
\setlength{\unitlength}{\pict@scale}
\hbox{\begin{picture}(\thepict@width,\thepict@height)
\put(0,\thepict@height){\psfig{figure=#1,width=#3,height=#2,clip=,angle=270}}
\SetScale{0.283466457}
\SetWidth{1.763889}
{#4}
\end{picture}}
}
\newcommand{\psfigrol}[4]{%
\setcounter{pict@width}{1*\ratio{#2+\pict@scale/2}{\pict@scale}}
\setcounter{pict@height}{1*\ratio{#3+\pict@scale/2}{\pict@scale}}
\setlength{\unitlength}{\pict@scale}
\hbox{\begin{picture}(\thepict@width,\thepict@height)
\put(0,0){\psfig{figure=#1,width=#3,height=#2,clip=,angle=90}}
\SetScale{0.283466457}
\SetWidth{1.763889}
{#4}
\end{picture}}
}
\catcode`\@=12 
\newlength\listtextwidth

\catcode`\@=11 
\newlength{\@tabfninsert}
\newlength{\@tabfnwidth}
\newcommand{\tabfootnote}[2]{%
  \setlength{\@tabfninsert}{0.8em}
  \setlength{\@tabfnwidth}{\textwidth}
  \addtolength{\@tabfnwidth}{-\@tabfninsert}
  \addtolength{\@tabfnwidth}{-0.4em}
  \noindent\makebox[\@tabfninsert][r]{\footnotesize$^{#1}$\hfil}\hfill%
  \parbox[t]{\@tabfnwidth}{\footnotesize #2\hfill}}
\catcode`\@=12 

\begin{document}
%

\titlehead{
  DESY 16-035 \\
  Accepted for publication in Physics Letters B \\
  March 2016
\vspace{4cm}
}

\zeustitle{%
Limits on the effective quark radius from
      inclusive \boldmath$ep$ scattering at HERA
}

\zeusauthor{ZEUS Collaboration}

\zeusdate{}

\maketitle

\begin{abstract}\noindent
The high-precision HERA data allows searches up to TeV scales for Beyond
the Standard Model contributions to electron--quark scattering.
Combined measurements of the inclusive deep inelastic cross sections in
neutral and charged current $ep$ scattering corresponding to 
a luminosity of around 1 fb$^{-1}$ have been used in this analysis.
A new approach to the beyond the Standard Model analysis of the inclusive
$ep$ data is presented;
simultaneous fits of parton distribution functions together with
contributions of ``new physics'' processes were performed.
Results are presented considering a finite radius of quarks
within the quark form-factor model.
The resulting \CL{95} upper limit on the effective quark radius is
$0.43\cdot 10^{-16}$~cm.
\end{abstract}

\thispagestyle{empty}
\cleardoublepage
%
%
\pagenumbering{roman}

\begin{center}
{                      \Large  The ZEUS Collaboration              }
\end{center}

{\small\raggedright


H.~Abramowicz$^{25, u}$, 
I.~Abt$^{20}$, 
L.~Adamczyk$^{8}$, 
M.~Adamus$^{31}$, 
S.~Antonelli$^{2}$, 
V.~Aushev$^{17}$, 
O.~Behnke$^{10}$, 
U.~Behrens$^{10}$, 
A.~Bertolin$^{22}$, 
S.~Bhadra$^{33}$
% ...
}
\pagebreak[4]

{
\footnotesize


{\setlength{\parskip}{0.4em}
\makebox[3ex]{$^{1}$}
\begin{minipage}[t]{14cm}
{\it INFN Bologna, Bologna, Italy}~$^{A}$

\end{minipage}

\makebox[3ex]{$^{2}$}
\begin{minipage}[t]{14cm}
{\it University and INFN Bologna, Bologna, Italy}~$^{A}$

\end{minipage}

\makebox[3ex]{$^{3}$}
\begin{minipage}[t]{14cm}
{\it Physikalisches Institut der Universit\"at Bonn,
Bonn, Germany}~$^{B}$

\end{minipage}

\makebox[3ex]{$^{4}$}
\begin{minipage}[t]{14cm}
{\it Panjab University, Department of Physics, Chandigarh, India}

\end{minipage}

\makebox[3ex]{$^{5}$}
\begin{minipage}[t]{14cm}
{\it Calabria University,
Physics Department and INFN, Cosenza, Italy}~$^{A}$

\end{minipage}

\makebox[3ex]{$^{6}$}
\begin{minipage}[t]{14cm}
{\it National Centre for Particle Physics, Universiti Malaya, 50603 Kuala Lumpur, Malaysia}~$^{C}$

\end{minipage}

\makebox[3ex]{$^{7}$}
\begin{minipage}[t]{14cm}
{\it The Henryk Niewodniczanski Institute of Nuclear Physics, Polish Academy of \\
Sciences, Krakow, Poland}~$^{D}$

\end{minipage}

\makebox[3ex]{$^{8}$}
\begin{minipage}[t]{14cm}
{\it AGH-University of Science and Technology, Faculty of Physics and Applied Computer
Science, Krakow, Poland}~$^{D}$

\end{minipage}

\makebox[3ex]{$^{9}$}
\begin{minipage}[t]{14cm}
{\it Department of Physics, Jagellonian University, Krakow, Poland}

\end{minipage}

\makebox[3ex]{$^{10}$}
\begin{minipage}[t]{14cm}
{\it Deutsches Elektronen-Synchrotron DESY, Hamburg, Germany}

\end{minipage}

\makebox[3ex]{$^{11}$}
\begin{minipage}[t]{14cm}
{\it Deutsches Elektronen-Synchrotron DESY, Zeuthen, Germany}

\end{minipage}

\makebox[3ex]{$^{12}$}
\begin{minipage}[t]{14cm}
{\it School of Physics and Astronomy, University of Glasgow,
Glasgow, United Kingdom}~$^{E}$

\end{minipage}

\makebox[3ex]{$^{13}$}
\begin{minipage}[t]{14cm}
{\it Hamburg University, Institute of Experimental Physics, Hamburg,
Germany}~$^{F}$

\end{minipage}

\makebox[3ex]{$^{14}$}
\begin{minipage}[t]{14cm}
{\it Institute of Particle and Nuclear Studies, KEK,
Tsukuba, Japan}~$^{G}$

\end{minipage}

\makebox[3ex]{$^{15}$}
\begin{minipage}[t]{14cm}
{\it Institute of Physics and Technology of Ministry of Education and
Science of Kazakhstan, Almaty, Kazakhstan}

\end{minipage}

\makebox[3ex]{$^{16}$}
\begin{minipage}[t]{14cm}
{\it Institute for Nuclear Research, National Academy of Sciences, Kyiv, Ukraine}

\end{minipage}

\makebox[3ex]{$^{17}$}
\begin{minipage}[t]{14cm}
{\it Department of Nuclear Physics, National Taras Shevchenko University of Kyiv, Kyiv, Ukraine}

\end{minipage}

\makebox[3ex]{$^{18}$}
\begin{minipage}[t]{14cm}
{\it Meiji Gakuin University, Faculty of General Education,
Yokohama, Japan}~$^{G}$

\end{minipage}

\makebox[3ex]{$^{19}$}
\begin{minipage}[t]{14cm}
{\it Lomonosov Moscow State University, Skobeltsyn Institute of Nuclear Physics,
Moscow, Russia}~$^{H}$

\end{minipage}

\makebox[3ex]{$^{20}$}
\begin{minipage}[t]{14cm}
{\it Max-Planck-Institut f\"ur Physik, M\"unchen, Germany}

\end{minipage}

\makebox[3ex]{$^{21}$}
\begin{minipage}[t]{14cm}
{\it Department of Physics, University of Oxford,
Oxford, United Kingdom}~$^{E}$

\end{minipage}

\makebox[3ex]{$^{22}$}
\begin{minipage}[t]{14cm}
{\it INFN Padova, Padova, Italy}~$^{A}$

\end{minipage}

\makebox[3ex]{$^{23}$}
\begin{minipage}[t]{14cm}
{\it Dipartimento di Fisica e Astronomia dell' Universit\`a and INFN,
Padova, Italy}~$^{A}$

\end{minipage}

\makebox[3ex]{$^{24}$}
\begin{minipage}[t]{14cm}
{\it Polytechnic University, Tokyo, Japan}~$^{G}$

\end{minipage}

\makebox[3ex]{$^{25}$}
\begin{minipage}[t]{14cm}
{\it Raymond and Beverly Sackler Faculty of Exact Sciences, School of Physics, \\
Tel Aviv University, Tel Aviv, Israel}~$^{I}$

\end{minipage}

\makebox[3ex]{$^{26}$}
\begin{minipage}[t]{14cm}
{\it Department of Physics, Tokyo Institute of Technology,
Tokyo, Japan}~$^{G}$

\end{minipage}

\makebox[3ex]{$^{27}$}
\begin{minipage}[t]{14cm}
{\it Universit\`a di Torino and INFN, Torino, Italy}~$^{A}$

\end{minipage}

\makebox[3ex]{$^{28}$}
\begin{minipage}[t]{14cm}
{\it Universit\`a del Piemonte Orientale, Novara, and INFN, Torino,
Italy}~$^{A}$

\end{minipage}

\makebox[3ex]{$^{29}$}
\begin{minipage}[t]{14cm}
{\it Physics and Astronomy Department, University College London,
London, United Kingdom}~$^{E}$

\end{minipage}

\makebox[3ex]{$^{30}$}
\begin{minipage}[t]{14cm}
{\it Faculty of Physics, University of Warsaw, Warsaw, Poland}

\end{minipage}

\makebox[3ex]{$^{31}$}
\begin{minipage}[t]{14cm}
{\it National Centre for Nuclear Research, Warsaw, Poland}

\end{minipage}

\makebox[3ex]{$^{32}$}
\begin{minipage}[t]{14cm}
{\it Department of Particle Physics and Astrophysics, Weizmann
Institute, Rehovot, Israel}

\end{minipage}

\makebox[3ex]{$^{33}$}
\begin{minipage}[t]{14cm}
{\it Department of Physics, York University, Ontario, Canada M3J 1P3}~$^{J}$

\end{minipage}

}

}

\pagebreak[4]

{
\small


{\setlength{\parskip}{0.4em}\raggedright
\makebox[3ex]{$^{ A}$}
\begin{minipage}[t]{14cm}
 supported by the Italian National Institute for Nuclear Physics (INFN) \
\end{minipage}

\makebox[3ex]{$^{ B}$}
\begin{minipage}[t]{14cm}
 supported by the German Federal Ministry for Education and Research (BMBF), under
 contract No. 05 H09PDF\
\end{minipage}

\makebox[3ex]{$^{ C}$}
\begin{minipage}[t]{14cm}
 supported by HIR grant UM.C/625/1/HIR/149 and UMRG grants RU006-2013, RP012A-13AFR and RP012B-13AFR from
 Universiti Malaya, and ERGS grant ER004-2012A from the Ministry of Education, Malaysia\
\end{minipage}

\makebox[3ex]{$^{ D}$}
\begin{minipage}[t]{14cm}
 supported by the National Science Centre under contract No. DEC-2012/06/M/ST2/00428\
\end{minipage}

\makebox[3ex]{$^{ E}$}
\begin{minipage}[t]{14cm}
 supported by the Science and Technology Facilities Council, UK\
\end{minipage}

\makebox[3ex]{$^{ F}$}
\begin{minipage}[t]{14cm}
 supported by the German Federal Ministry for Education and Research (BMBF), under
 contract No. 05h09GUF, and the SFB 676 of the Deutsche Forschungsgemeinschaft (DFG) \
\end{minipage}

\makebox[3ex]{$^{ G}$}
\begin{minipage}[t]{14cm}
 supported by the Japanese Ministry of Education, Culture, Sports, Science and Technology
 (MEXT) and its grants for Scientific Research\
\end{minipage}

\makebox[3ex]{$^{ H}$}
\begin{minipage}[t]{14cm}
 supported by RF Presidential grant N 3042.2014.2 for the Leading Scientific Schools\
\end{minipage}

\makebox[3ex]{$^{ I}$}
\begin{minipage}[t]{14cm}
 supported by the Israel Science Foundation\
\end{minipage}

\makebox[3ex]{$^{ J}$}
\begin{minipage}[t]{14cm}
 supported by the Natural Sciences and Engineering Research Council of Canada (NSERC) \
\end{minipage}

}

\pagebreak[4]
{\setlength{\parskip}{0.4em}


\makebox[3ex]{$^{ a}$}
\begin{minipage}[t]{14cm}
now at INFN Roma, Italy\
\end{minipage}

\makebox[3ex]{$^{ b}$}
\begin{minipage}[t]{14cm}
now at Sant Longowal Institute of Engineering and Technology, Longowal, Punjab, India\
\end{minipage}

\makebox[3ex]{$^{ c}$}
\begin{minipage}[t]{14cm}
now at Sri Guru Granth Sahib World University, Fatehgarh Sahib, India\
\end{minipage}

\makebox[3ex]{$^{ d}$}
\begin{minipage}[t]{14cm}
also at Agensi Nuklear Malaysia, 43000 Kajang, Bangi, Malaysia\
\end{minipage}

\makebox[3ex]{$^{ e}$}
\begin{minipage}[t]{14cm}
partially supported by the Polish National Science Centre projects DEC-2011/01/B/ST2/03643 and DEC-2011/03/B/ST2/00220\
\end{minipage}

\makebox[3ex]{$^{ f}$}
\begin{minipage}[t]{14cm}
now at Rockefeller University, New York, NY 10065, USA\
\end{minipage}

\makebox[3ex]{$^{ g}$}
\begin{minipage}[t]{14cm}
now at European X-ray Free-Electron Laser facility GmbH, Hamburg, Germany\
\end{minipage}

\makebox[3ex]{$^{ h}$}
\begin{minipage}[t]{14cm}
now at University of Liverpool, United Kingdom\
\end{minipage}

\makebox[3ex]{$^{ i}$}
\begin{minipage}[t]{14cm}
now at Tel Aviv University, Isreal\
\end{minipage}

\makebox[3ex]{$^{ j}$}
\begin{minipage}[t]{14cm}
now at Physikalisches Institut, Universit\"{a}t Heidelberg, Germany\
\end{minipage}

\makebox[3ex]{$^{ k}$}
\begin{minipage}[t]{14cm}
supported by the Alexander von Humboldt Foundation\
\end{minipage}

\makebox[3ex]{$^{ l}$}
\begin{minipage}[t]{14cm}
now at CERN, Geneva, Switzerland\
\end{minipage}

\makebox[3ex]{$^{ m}$}
\begin{minipage}[t]{14cm}
Alexander von Humboldt Professor; also at DESY and University of Oxford\
\end{minipage}

\makebox[3ex]{$^{ n}$}
\begin{minipage}[t]{14cm}
also at DESY\
\end{minipage}

\makebox[3ex]{$^{ o}$}
\begin{minipage}[t]{14cm}
also at University of Tokyo, Japan\
\end{minipage}

\makebox[3ex]{$^{ p}$}
\begin{minipage}[t]{14cm}
now at Kobe University, Japan\
\end{minipage}

\makebox[3ex]{$^{ q}$}
\begin{minipage}[t]{14cm}
member of National Technical University of Ukraine, Kyiv Polytechnic Institute, Kyiv, Ukraine\
\end{minipage}

\makebox[3ex]{$^{ r}$}
\begin{minipage}[t]{14cm}
now at DESY CMS group\
\end{minipage}

\makebox[3ex]{$^{ s}$}
\begin{minipage}[t]{14cm}
now at DESY ATLAS group\
\end{minipage}

\makebox[3ex]{$^{ t}$}
\begin{minipage}[t]{14cm}
now at LNF, Frascati, Italy\
\end{minipage}

\makebox[3ex]{$^{ u}$}
\begin{minipage}[t]{14cm}
also at Max Planck Institute for Physics, Munich, Germany, External Scientific Member\
\end{minipage}

\makebox[3ex]{$^{ v}$}
\begin{minipage}[t]{14cm}
also supported by DESY and the Alexander von Humboldt Foundation\
\end{minipage}

\makebox[3ex]{$^{ w}$}
\begin{minipage}[t]{14cm}
also at \L\'{o}d\'{z} University, Poland\
\end{minipage}

\makebox[3ex]{$^{ x}$}
\begin{minipage}[t]{14cm}
member of \L\'{o}d\'{z} University, Poland\
\end{minipage}

\makebox[3ex]{$^{ y}$}
\begin{minipage}[t]{14cm}
now at Polish Air Force Academy in Deblin\
\end{minipage}

}

}

\newpage


\pagenumbering{arabic}

\section{Introduction}
\label{sec-int}

Precision measurements of deep inelastic $e^\pm p$ scattering (DIS)
cross sections at high values of negative
four-momentum-transfer squared,  $Q^2$, 
allow searches for contributions beyond the Standard Model (BSM),
even far beyond the centre-of-mass energy of the $e^\pm p$ interactions.
For many ``new physics'' scenarios, cross sections can be affected
by new kinds of interactions in which virtual BSM particles are exchanged.
The cross sections would also be influenced were quarks to have a finite radius. 
As the HERA kinematic range is assumed to be far below the scale of
the new physics, all such BSM interactions can be approximated as
contact interactions (CI). 
In all cases,  deviations of the observed cross section 
from the Standard Model (SM) prediction  are searched for 
in $ep$ scattering at the highest available $Q^2$.
The predictions are calculated using parton distribution function (PDF)
parameterisations of the proton.

The H1 and ZEUS collaborations measured inclusive 
$e^{\pm}p$ scattering cross sections at HERA from
1994 to 2000 (HERA I) and from 2002 to 2007 (HERA II), 
collecting together a total integrated luminosity of about 1\,fb$^{-1}$.
All inclusive data were recently combined \cite{h1zeus_inc} to create
one consistent set of neutral current (NC) and charged current (CC)
cross-section measurements for $e^{\pm}p$ scattering with unpolarised beams.
The inclusive cross sections were used as input to a QCD analysis
within the DGLAP formalism, resulting in a PDF set
denoted as \mbox{HERAPDF2.0}.
Due to the high precision and consistency of the input data,
HERAPDF2.0 can be used to calculate SM predictions with small 
uncertainties.
A search for BSM contributions in the data should take into account the
possibility that the PDF set may already have been biased by partially
or totally absorbing previously unrecognised BSM contributions.

In the ZEUS CI analysis of HERA I  $e^{\pm}p$ data~\cite{pl:b591:23}, 
the uncertainties on the PDFs used were a dominant source of systematic error.
Estimated uncertainties of the parton densities were used to smear model
predictions in the limit-setting procedure.
Such an approach was valid as the CTEQ5D parameterisation 
\cite{epj:c12:375, pr:d55:1280} used for calculating model predictions 
included only 1994 HERA data in addition to many other data sets.
The limits were dominated by statistical uncertainties.
For the CI analysis presented here, in which the data are identical to
those used for the \mbox{HERAPDF2.0} determination and the statistical
uncertainties are no longer dominant,
a new procedure to set limits on the BSM model contributions is required.
In this analysis BSM contributions and the QCD evolution are
fitted simultaneously.
Results of a search for a finite quark radius are presented within
the formalism of the quark form-factor model~\cite{Kopp:1994qv}.


\section{QCD analysis}
\label{sec-qcd}

The QCD analysis presented in this paper was performed similarly to
that for the HERAPDF2.0 determination~\cite{h1zeus_inc}.
It was used to predict cross sections without BSM contributions.
The HERA combined data on inclusive $e^\pm p$ scattering~\cite{h1zeus_inc} 
were used as input to the perturbative QCD (pQCD) analysis.
Only cross sections with $Q^2 \geqslant 3.5$\,GeV$^2$ were used.
A fit to the data, resulting in a set of PDFs, was obtained by solving 
the DGLAP evolution equations at NLO in the \mbox{$\overline{\rm{MS}}$} scheme.
This was done using the programme QCDNUM~\cite{QCDNUM} within the HERAFitter 
framework~\cite{HERAfitter}. 
For the PDF parameterisation, the approach adopted in the \mbox{HERAPDF2.0} 
study~\cite{h1zeus_inc} was followed. 
The PDFs of the proton were described at a starting scale of $1.9$ GeV$^2$ 
in terms of 14 parameters.
These parameters were fit to the data using a $\chi^2$ method, 
taking into account statistical uncertainties, as well as 
uncorrelated and correlated systematic uncertainties on the
experimental data.
The corresponding $\chi^2$ formula is:
\begin{equation}
 \chi^2 \left(\boldsymbol{m},\boldsymbol{s}\right) = 
 \sum_i
 \frac{\left[m^i
+ \sum_j \gamma^i_j m^i s_j  - {\mu_{0}^i} \right]^2}
{\left( \textstyle \delta^2_{i,{\rm stat}} +
\delta^2_{i,{\rm uncor}} \right) \,  (\mu_{0}^i)^2}
 + \sum_j s^2_j ~~,
\label{eq:qcdfit}
\end{equation} 
where $\mu_{0}^{i}$ is the measured cross-section value at the point $i$.
The quantities $\gamma^{i}_j $, $\delta_{i,{\rm stat}} $ and 
$\delta_{i,{\rm uncor}}$ are the relative correlated 
systematic, relative statistical and relative uncorrelated 
systematic uncertainties of the input data, respectively. 
The vector $\boldsymbol{m}$ represents the set of pQCD cross-section 
predictions $m^i$ and the components $s_j$ of the vector
$\boldsymbol{s}$ represent the correlated systematic shifts of the
cross sections (given in units of $\gamma^{i}_j $).
The summations extend over all data points $i$ and all correlated 
systematic uncertainties $j$.

The $\chi^2$ formula used  in this analysis differs from that of 
\mbox{HERAPDF2.0} study~\cite{h1zeus_inc}
in order to facilitate the production of data replicas within the HERAFitter 
framework~\cite{HERAfitter}, see Section~\ref{sec-limit}.
The resulting sets of PDFs, referred to as ZRqPDF in the following,
are nevertheless in good agreement with \mbox{HERAPDF2.0}.

The experimental uncertainties on the predictions from ZRqPDF were 
determined with the criterion $\Delta\chi^2=1$. 
The uncertainties due to the choice of model settings and the 
form of the parameterisation were evaluated as for \mbox{HERAPDF2.0}.


\section{Quark form factor}
\label{sec-rq}

One of the possible parameterisations of deviations from SM predictions
in $ep$ scattering is achieved by assigning an effective finite radius to 
electrons and/or quarks while assuming the SM gauge bosons remain 
point-like and their couplings unchanged.
The expected modification of the SM cross section can be described using
a semi-classical form-factor approach~\cite{Kopp:1994qv}.
If the expected deviations are small, the SM predictions for the 
cross sections are modified, approximately, to:
\begin{eqnarray}
\frac{d\sigma}{dQ^{2}} & = & 
\frac{d\sigma^{\rm SM}}{dQ^{2}} \;
\left( 1 - \frac{R_{e}^{2}}{6} \, Q^{2} \right)^{2} \;
\left( 1 - \frac{R_{q}^{2}}{6} \, Q^{2} \right)^{2} \; ,   \label{eq:rq}
\end{eqnarray}
where $R_{e}^{2}$ and $R_{q}^{2}$ are the mean-square radii of the 
electron and the quark, respectively,
related to new BSM energy scales.
In the present analysis, only the possible finite spatial distribution 
of the quark was considered and the electron was assumed to be point-like 
($R_{e}^{2} \equiv 0$). 
Both positive and negative values of $R_{q}^{2}$ were considered.
Negative values of $R_{q}^{2}$ can be obtained if a charge distribution 
is assumed which changes sign as a function of the radius.
The term ``quark radius'' is only one possible interpretation of
BSM effects parameterised as form factors.

The QCD analysis described in the previous section was extended
by introducing $R_{q}^{2}$ as an additional model parameter and modifying all
$e^{\pm} p$ DIS cross-section predictions according to Eq.~\ref{eq:rq}.
Values for $R_{q}^{2}$ were extracted using a $\chi^2$-minimisation 
procedure, where  all PDF parameters were also simultaneously fit;
$R_{q}^{2}$ was treated as a test statistic to be used for limit setting.
The value of this test statistic for the data is
\(R_q^{\rm 2\; Data} = -0.2\cdot 10^{-33}\; {\rm cm}^2\).
The probability distributions for $R_q^2$ were determined as described in 
the next section.


\section{Limit-setting procedure}
\label{sec-limit}

The limit on the effective quark-radius squared, $R_{q}^{2}$, is derived
in a frequentist approach \cite{Cousins:1994yw} using the technique of
replicas.  
Replicas are sets of cross-section values that are
generated by varying all cross sections randomly according to their
known uncertainties.
For the analysis presented here, multiple replica sets were used, each 
covering cross-section values on all points of the $x,Q^2$ grid
used in the QCD fit.
For an assumed true value of the quark-radius squared,
$R_q^{\rm 2\;True}$, replica data sets were created by taking
the reduced cross sections calculated from the ZRqPDF fit and scaling them
with the quark form factor, Eq.~\ref{eq:rq}, with 
$R_{q}^{2} = R_q^{\rm 2\;True}$.
This results in a set of cross-section values $m_{0}^{i}$ for the assumed
true quark-radius squared, $R_q^{\rm 2\;True}$.
The values of $m_{0}^{i}$ were then varied randomly within statistical
and systematic uncertainties taken from the data, taking correlations
into account.
All uncertainties were assumed to follow a Gaussian
distribution\footnote{It was verified that using a Poisson probability
distribution for producing replicas at high $Q^2$, where the event
samples are small, and using the $\chi^2$ minimisation for
these data did not significantly change the probability distributions for
the fitted parameter values.}.
For each replica, the generated value of the cross section 
at the point $i$, $\mu^{i}$, was calculated as:
\begin{equation}
 \mu^{i}  = 
 \left[ m_{0}^{i} + \sqrt{\delta^2_{i,{\rm stat}} + \delta^2_{i, {\rm uncor}}} \cdot  \mu_{0}^{i} \cdot r_i \right]
\cdot
\left( 1 + \sum_j \gamma^i_j \cdot r_j  \right)~~,
\label{eq:replica}
\end{equation} 
where variables $r_i$ and $r_j$ represent random numbers from a normal
distribution for each data point $i$ and for each source of correlated
systematic uncertainty $j$, respectively.

The approach adopted was to generate sets of replicas that were used
to test the hypothesis that the cross sections were modified by a fixed
$R_{q}^{2}$ value according to Eq.~\ref{eq:rq}.
The value of $R_q^{\rm 2\;Data}$  determined by the fit to the data themselves
was taken as a test statistic, to which values from 
fits to replicas, $R_q^{\rm 2\;Fit}$, could be compared.
Positive (negative) $R_{q}^{\rm 2\;True}$  values that,
in more than 95\% of the replicas, result in the fitted radius squared value,
$R_{q}^{\rm 2\;Fit}$, greater than (less than) that obtained for the data,
$R_q^{\rm 2\; Data}$, were excluded at the \CL{95}.
The details of these procedures are described below.

To set the limit, a number of MC replica cross-section sets for each value 
of $R_{q}^{\rm 2\;True}$ was used for a QCD fit with the PDF parameters and 
the quark radius as free parameters, yielding a distribution of the fitted 
values of the quark radius, $R_{q}^{\rm 2 \; Fit}$.
The $\chi^2$ formula of Eq.~\ref{eq:qcdfit},
with the measured cross-section values,
$\mu_{0}^{i}$, in the numerator of the first term replaced by the
generated values of the replica, $\mu^{i}$,
was used for fitting $R_{q}^{2}$ and the PDF parameters.

\begin{figure}[tbp]
\vfill
\begin{center}
\includegraphics[width=0.9\textwidth]{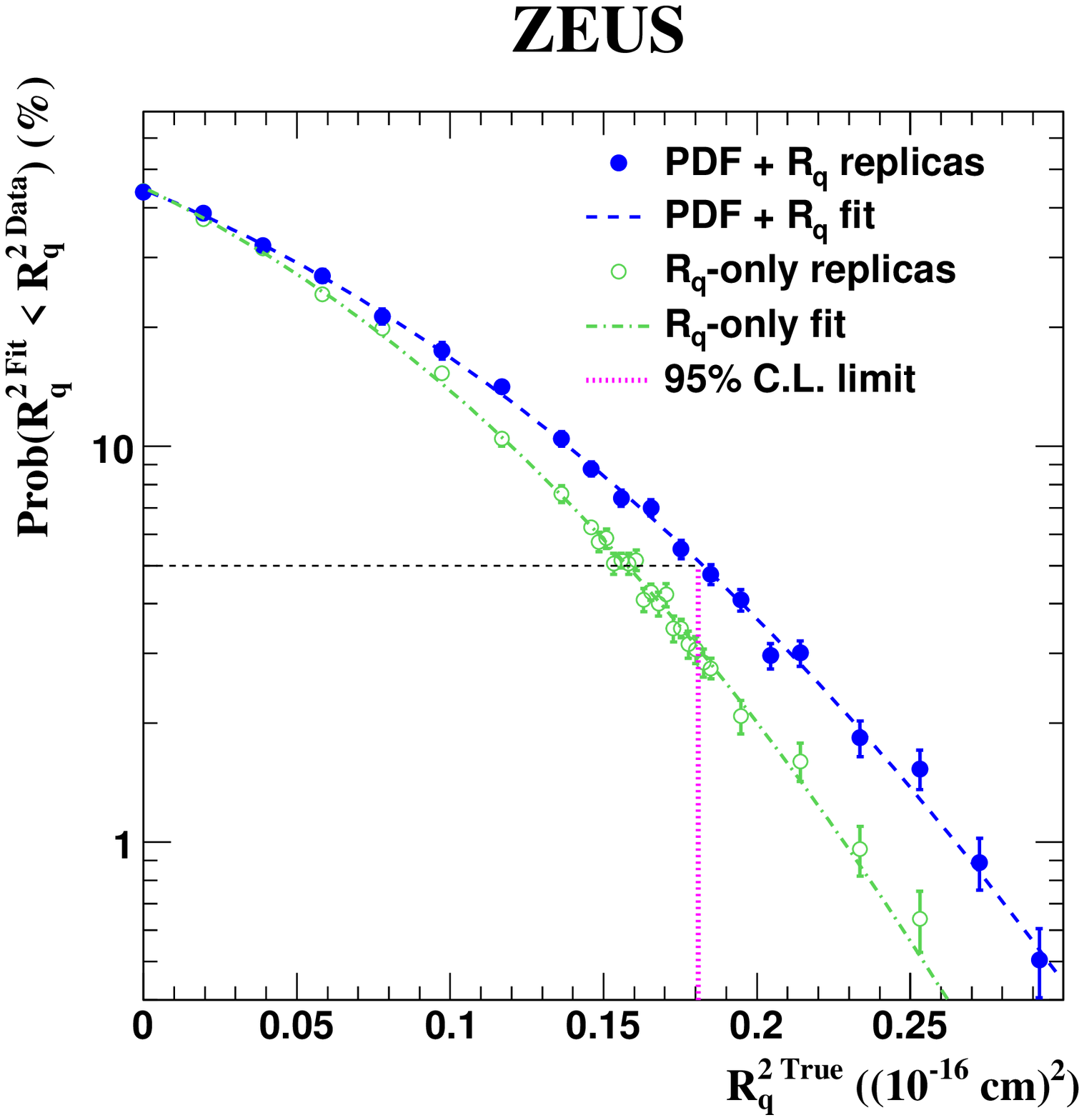}
\end{center}
\caption{
The probability of obtaining $R_{q}^{2 \; Fit}$ values smaller 
than that obtained for the actual data, $R_q^{2\; Data}$, calculated
from Monte Carlo replicas, as a function of the assumed value 
for the quark-radius squared, $R_{q}^{2\;True}$. 
Points with statistical error bars represent Monte Carlo replica sets
generated for different values of $R_{q}^{2\;True}$.
The solid circles correspond to the results obtained from the
simultaneous fit of $R_{q}^{2}$ and PDF parameters (PDF+$R_q$).
For comparison, the open circles represent the dependence obtained 
when fixing the PDF parameters to the ZRqPDF values ($R_q$-only).
The dashed line and the dashed--dotted line 
represent the cumulative Gaussian distributions fitted
to the PDF+$R_q$ and $R_q$-only replica points, respectively.
The vertical line represents the \CL{95} upper limit on
$R_{q}^{2}$.
}
\label{fig-rq_prob_central}
\vfill
\end{figure}

\begin{figure}[tbp]
\vfill
\begin{center}
\includegraphics[width=0.9\textwidth]{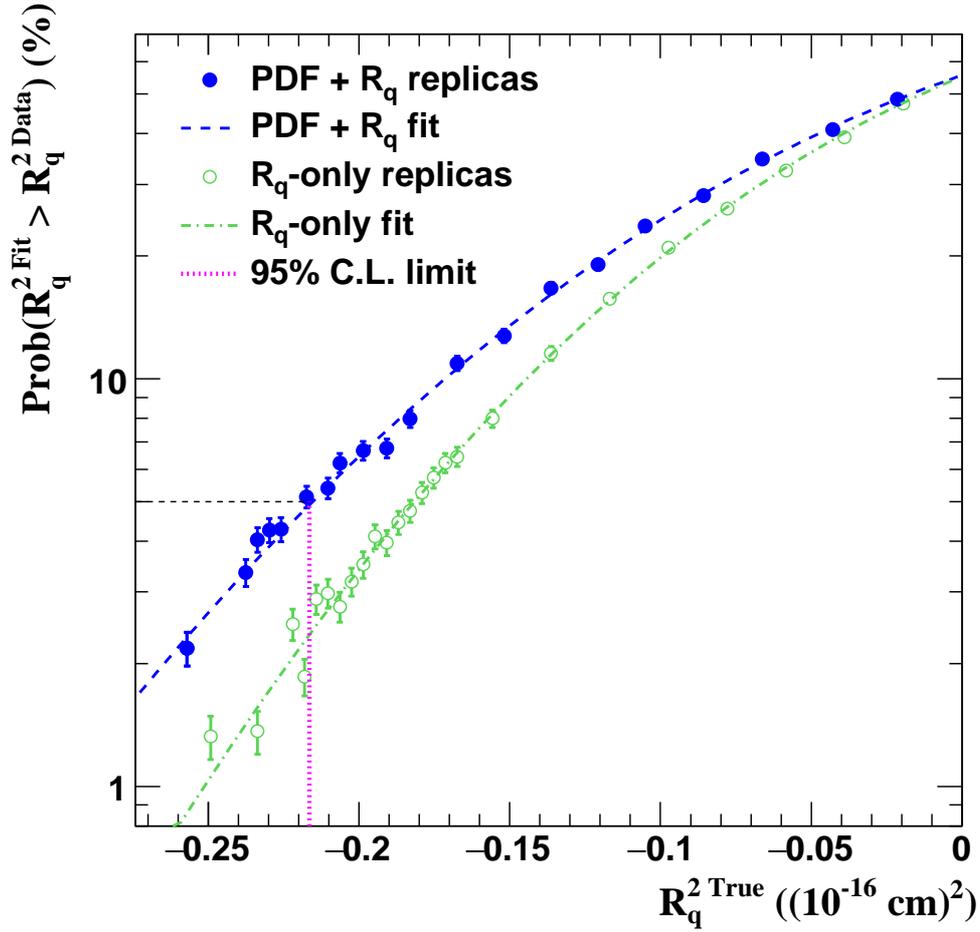}
\end{center}
\caption{
The probability of obtaining $R_{q}^{2 \; Fit}$ values larger
than that obtained for the actual data, $R_q^{2\; Data}$, calculated
from Monte Carlo replicas, as a function of the assumed value 
for the quark-radius squared, $R_{q}^{2\;True}$. 
Other details as for \fig{rq_prob_central}.
%
%
%
%
%
}
\label{fig-rq_prob_neg}
\vfill
\end{figure}

In a last step, the probability of obtaining a $R_{q}^{\rm 2 \; Fit}$ value
smaller than that obtained for the actual data, 
$\text{Prob}(R_{q}^{\rm 2 \; Fit} < R_q^{\rm 2\; Data})$,
was plotted as a function of $R_{q}^{\rm 2\; True}$,
for positive $R_{q}^{\rm 2\; True}$ values,
as shown in \fig{rq_prob_central}.
The probability distribution was interpolated to calculate
the $R_{q}^{2}$ value corresponding to the \CL{95} upper limit.
About 5000 Monte Carlo replicas were generated for each value
of $R_{q}^{\rm 2\; True}$ resulting in a relative statistical uncertainty
of the extracted limit of about 0.3\%.
The corresponding plot for negative $R_{q}^{\rm 2\; True}$ values is shown 
in \fig{rq_prob_neg}.

As a cross check, the limits on $R_{q}^{2}$ were also estimated from the 
simultaneous PDF and  $R_{q}^{2}$ fit to the data by looking at the
variation of the $\chi^2$ value minimised with respect to the PDF
parameters when changing the  $R_{q}^{2}$ value. 
Both limits are in good agreement with the results based on
the Monte Carlo replicas. 
The limit-setting procedure was also repeated for different
model and parameter settings, considered as systematic checks
in the \mbox{HERAPDF2.0} analysis~\cite{h1zeus_inc}.
The resulting variations of the limits on $R_{q}^{2}$ are negligible.
%


\section{Results}
\label{sec-results}

\begin{figure}[tbp]
\vfill
\begin{center}
\includegraphics[width=0.9\textwidth]{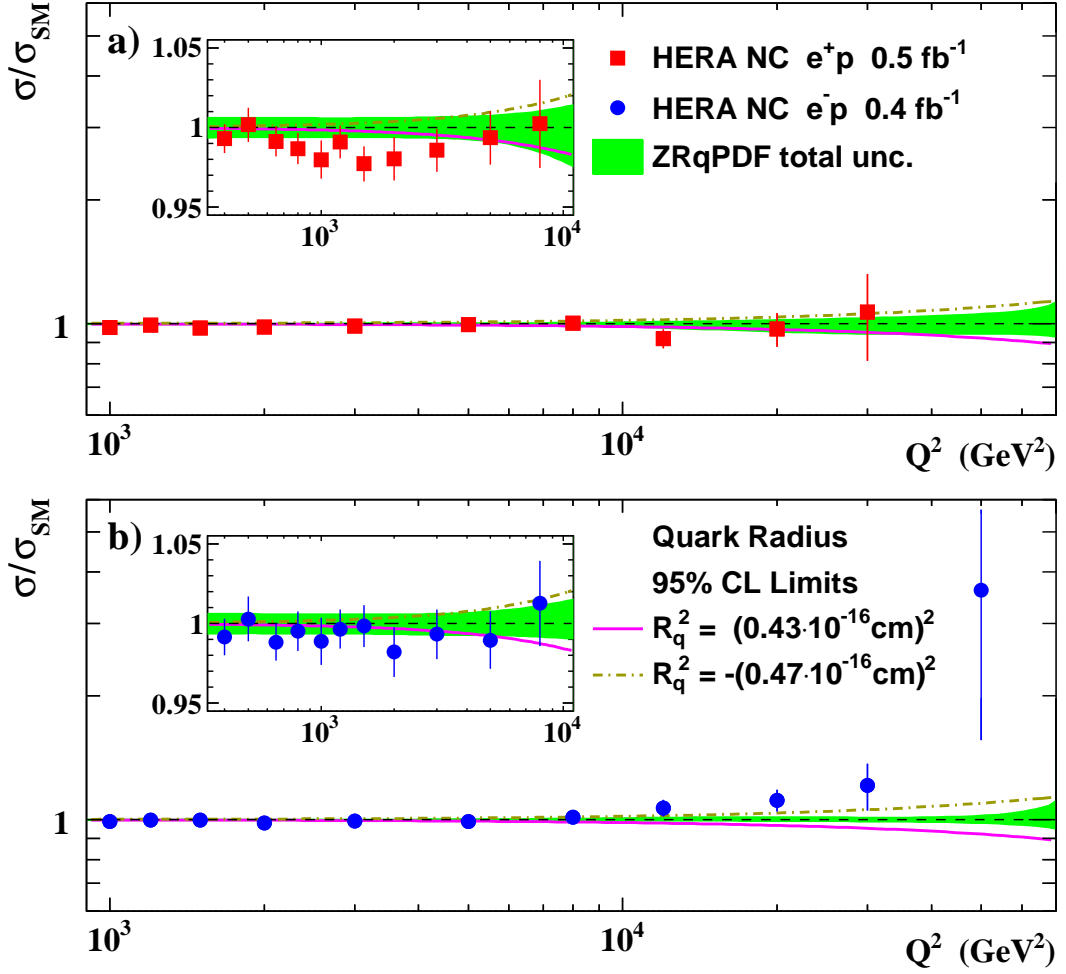}
\end{center}
\caption{
Combined HERA (a) $e^+p$ and (b) $e^-p$ NC DIS data 
compared to the \CL{95} exclusion limits on 
the effective mean-square radius of quarks. 
Also shown are the expectations 
calculated using the ZRqPDF parton distributions. 
The bands represent the total uncertainty on the predictions. 
The insets show the comparison in the $Q^{2} < 10^{4}$~GeV$ \, {}^{2}$ 
region with a linear ordinate scale.
}
\label{fig-rq_limit}
\vfill
\end{figure}

\begin{figure}[tbp]
\vfill
\begin{center}
\includegraphics[width=0.9\textwidth]{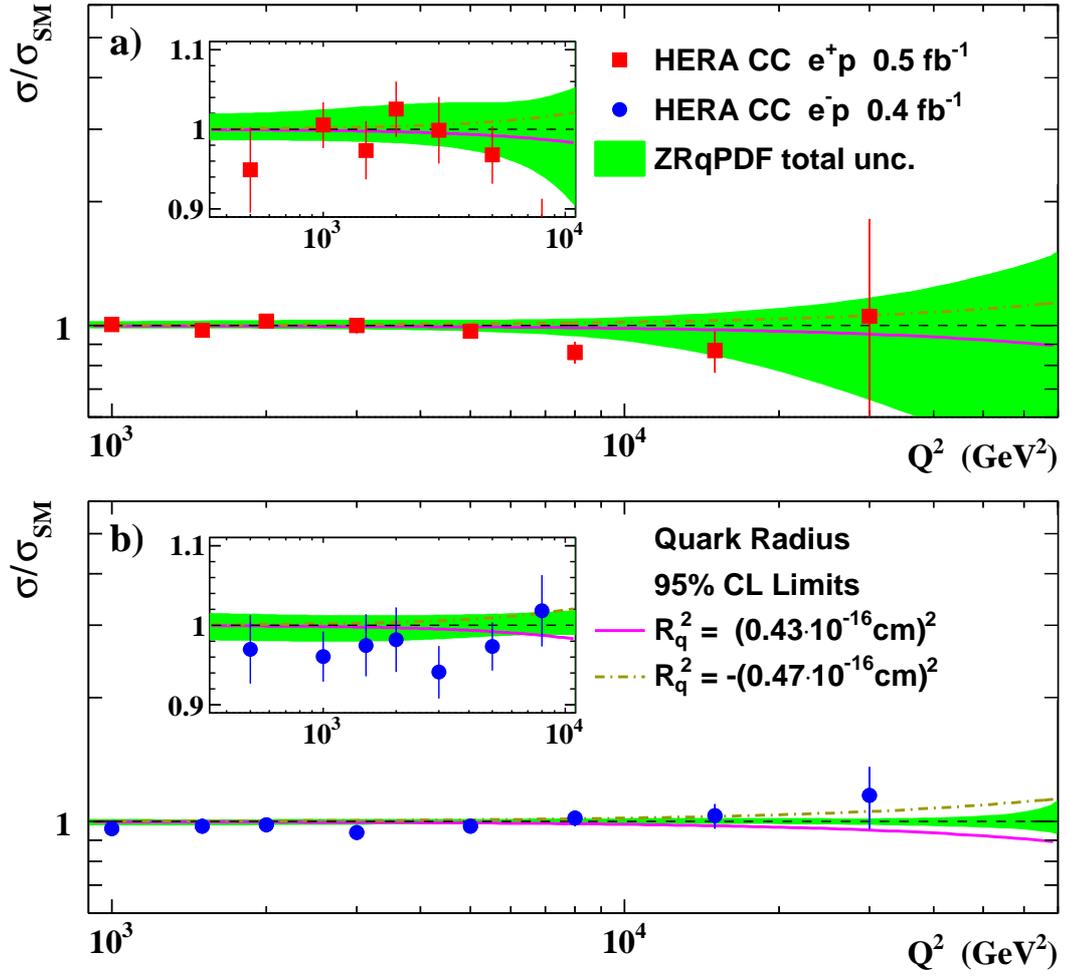}
\end{center}
\caption{
Combined HERA (a) $e^+p$ and (b) $e^-p$  CC DIS data 
compared to the \CL{95} exclusion limits on 
the effective mean-square radius of quarks. 
Other details as for \fig{rq_limit}.
}
\label{fig-rq_limit_cc}
\vfill
\end{figure}

The results of the limit-setting procedure using the simultaneous fit to PDF
parameters and $R_{q}^{2}$, based on sets of Monte
Carlo replicas testing the possible cross-section modifications
due to a quark form factor, yield the \CL{95} limits on the effective 
quark radius of
\begin{eqnarray*}
  -(0.47\cdot 10^{-16} \cm)^2 \; < \;
     R_q^{2} & < & (0.43\cdot 10^{-16} \cm)^2 \; .
\end{eqnarray*}  
Taking into account the possible influence of quark radii on the PDF 
parameters is necessary as demonstrated in 
Figs.~\ref{fig-rq_prob_central}~and~\ref{fig-rq_prob_neg}, because the 
limits that would be obtained for fixed PDF parameters are too strong
by about 10\%.
The limits are consistent with the estimated experimental sensitivity,
calculated as the median of the limit distribution
for the SM replicas, corresponding to a quark radius of
$0.45\cdot 10^{-16} \cm$ (for both positive and negative $R_{q}^{2}$).
Cross-section deviations given by Eq.~\ref{eq:rq}, corresponding
to the presented \CL{95} exclusion limits,
are compared to the combined HERA high-$Q^{2}$ NC and CC DIS data
in Figs.~\ref{fig-rq_limit} and \ref{fig-rq_limit_cc}, respectively.

The \CL{95} upper limit for the quark radius presented here is almost
a factor of two better than the previous ZEUS limit of
\mbox{$0.85\cdot 10^{-16}$~cm}, based on the HERA~I data \cite{pl:b591:23}.
The present result improves the limit set in $ep$ scattering by the 
H1 collaboration~\cite{pl:b705:52} ($R_q < 0.65\cdot 10^{-16} \cm$) 
and is similar to the limit presented by the L3 collaboration 
($R_q < 0.42\cdot 10^{-16}\cm$), based on 
quark-pair production at LEP2 \cite{pl:b489:81}.
It is important to remember that the possible BSM physics parameterised 
by the $R_q$ at LEP and HERA can be very different, so that the LEP and 
HERA limits are largely complementary.
The limit on negative $R_{q}^{2}$ values presented here is an improvement
compared to the published ZEUS limit of
$R_q^{2} \; > \; -(1.06\cdot 10^{-16} \cm)^2$.
%


\section{Conclusions}
\label{sec-conclusions}

The HERA combined measurement of inclusive deep inelastic cross
sections in neutral and charged current $e^{\pm} p$ scattering was used 
to set limits on possible deviations from the Standard Model due to a finite 
radius of the quarks.
The limit-setting procedure was based on a simultaneous fit of PDF 
parameters and the quark radius.
The resulting \CL{95} limits for the quark radius are 
\begin{eqnarray*}
  -(0.47\cdot 10^{-16} \cm)^2 \; < \;
     R_q^{2} & < & (0.43\cdot 10^{-16} \cm)^2 \; .
\end{eqnarray*}
This result is competitive with a determination from LEP2 and substantially
improves previous HERA limits.
%


\section*{Acknowledgements}
\label{sec-acknowledgements}

\Zacknowledge


{
  \bibliographystyle{l4z_pl3}
{\raggedright
  \bibliography{DESY-16-035.bib}}
}
\vfill\eject



\end{document}